%% file: main.tex
\documentclass{aastex631}

\usepackage{aas_macros}
\usepackage{amsmath}
\usepackage{graphicx}
\usepackage{hyperref}
\usepackage{xcolor}

\begin{document}

\title{Analysis methods to localize and characterize X-ray sources with the Micro-channel X-ray Telescope on board the SVOM satellite}

\author{Shaymaa Hussein}
\affiliation{Universit\'e Paris-Saclay, CNRS/IN2P3, IJCLab, 91405 Orsay, France}
\affiliation{An-Najah National University, Nablus, Palestine}

\author{Florent Robinet}
\affiliation{Universit\'e Paris-Saclay, CNRS/IN2P3, IJCLab, 91405 Orsay, France}

\author{Martin Boutelier}
\affiliation{Centre National d’Etudes Spatiales, Centre spatial de Toulouse, 18 avenue Edouard Belin, 31401 Toulouse Cedex 9, France}

\author{Diego G\"otz}
\affiliation{Université Paris-Saclay, Université Paris Cité, CEA, CNRS, AIM, 91191, Gif-sur-Yvette, France}

\author{Aleksandra Gros}
\affiliation{Université Paris-Saclay, Université Paris Cité, CEA, CNRS, AIM, 91191, Gif-sur-Yvette, France}

\author{Benjamin Schneider}
\affiliation{Université Paris-Saclay, Université Paris Cité, CEA, CNRS, AIM, 91191, Gif-sur-Yvette, France}

\begin{abstract}
  SVOM is a Sino-French space mission targeting high-energy transient astrophysical objects such as gamma-ray bursts. The soft X-ray part of the spectrum is covered by the Micro-channel X-ray Telescope (MXT) which is a narrow field telescope designed to precisely localize X-ray sources. This paper presents the method implemented on-board to characterize and localize X-ray sources with the MXT.  A specific localization method was developed to accommodate the optical system of the MXT, which is based on ``Lobster-Eye'' grazing incidence micro-pore optics. For the first time, the algorithm takes advantage of cross-correlation techniques to achieve a localization accuracy down to 2~arcmin with less than $200$ photons, which guarantees a rapid follow-up for most of the gamma-ray bursts that SVOM will observe. In this paper, we also study the limitations of the algorithm and characterize its performance.
\end{abstract}

\input{Introduction}

\input{Input}
\input{Algorithm}
\input{Characterization}
\input{Conclusion}

\begin{acknowledgments}
This work was developed in the context of the SVOM/MXT science group. We gratefully acknowledge the support from the group and we value the fruitful discussions that took place. In particular, we are grateful to the MXT team at the University of Leicester for providing the flight-model point spread function used in this paper, with special thanks to Richard ``PSF wizard'' Willingale. We also thank the MXT team at the Max Planck Institute for Extraterrestrial Physics for hosting the final performance tests of the MXT conducted at the Panter facility, Neuried Germany, in November 2021. The PhD thesis of Shaymaa Hussein was supported by the Centre National d’Etudes Spatiales in France.
\end{acknowledgments}

% ---- Bibliography ----
\bibliographystyle{aasjournal}
\bibliography{biblio}

\end{document}

%% file: Introduction.tex
\section{Introduction}\label{sec:introduction}
The SVOM (Space-based multi-band astronomical Variable Object Monitor) is a space mission~\citep{Wei:2016eox} coordinated by the Chinese (CNSA) and French (CNES) space agencies. The core program of SVOM is to detect and characterize gamma ray bursts~\citep{zhang_2018}, which are intense flashes of gamma rays lasting from a fraction to a few hundreds of seconds. Additionnally, SVOM will also target other astrophysical objects: active galactic nuclei~\citep{doi:https://doi.org/10.1002/9783527666829.ch3}, black-hole binaries~\citep{McClintock:2003gx}, galactic X-ray binaries and magnetars~\citep{Revnivtsev:2008sn}, flaring stars~\citep{Kenyon:1987js}, or the cataclysmic variables~\citep{Robinson:1976bu}. Finally, SVOM will participate to multi-messenger programs where transient sources observed by other experiments (gravitational-wave detectors~\citep{KAGRA:2013rdx} or neutrino detectors) will be followed-up~\citep{zhang_2018}.

The SVOM satellite will be launched in late 2023. It is equipped with four instruments. At high energies, the ECLAIRs~\citep{Godet:2014ava} telescope ($>4$~keV) and the Gamma-Ray Monitor~\citep{Dong:2009xs} ($>30$~keV) will trigger on gamma-ray sources. These sources will then be followed-up by the Microchannel X-ray Telescope (MXT)~\citep{MXTGotz:2015mha} and the Visible Telescope (VT)~\citep{Wang:2009pn}. The MXT is designed to observe X-ray sources from 0.2 to 10 keV. For gamma-ray bursts, this energy range covers the afterglow emission~\citep{Costa:1997obd} following the gamma-ray burst prompt emission. The original trigger is either detected on-board by the ECLAIRs/Gamma-Ray Monitor or by another experiment. The SVOM platform then slews to place the astrophysical object in the MXT $58\times 58$~arcmin$^2$ field of view. When stable, the telescope cumulates X-ray photons. The MXT images are analyzed on-board to localize the X-ray source. The reconstructed position is updated every 30 seconds and transmitted to the ground and broadcasted world-wide to enable a follow-up of the event with ground telescopes. It will also be used to trigger a second slew and better position the source for an optical follow-up with the VT. One of the MXT scientific requirements is to detect and localize 85\% of gamma-ray bursts with an accuracy better than 2~arcmin after 10 minutes of stable observation. 

The MXT is a light-weight ($<42$~kg) and compact (focal length $\sim 1.15$~m) X-ray telescope. Its unique optical system is based on a ``Lobster-Eye'' grazing incidence technique~\citep{10.1117/12.2561739}. High-energy photons are reflected and focused by a collection of square micropores creating a unique pattern on the focal plane, composed of a central peak and two cross arms. The X-ray source sky location is derived from the position of the central peak in the focal plane.

This paper presents the algorithm developed to perform the on-board localization analysis. A cross-correlation technique is used to best resolve sources of X-ray photons, even with a weak intensity. This method is applied to search for multiple sources in the MXT field of view using an iterative subtraction process. For each identified source, the number of photons is evaluated using the cross-correlation data, as well as the overall background counts. This is a new approach which was specifically designed to manage the peculiar shape of the point spread function of the telescope. Moreover, due to space constraints, the data-processing methods had to be optimized to cope with low on-board computing resources. This method is characterized and we demonstrate that the scientific requirements of the MXT are achieved by the on-board software.

This paper is organized as follows. Section~\ref{sec:input} introduces the two data products needed by the localization software: the camera images and the telescope point spread function. Section~\ref{sec:algorithm} presents the on-board analysis steps: the source localization, the estimation of signal and background counts, and the method to identify multiple X-ray sources. This method is characterized in Sec.~\ref{sec:characterization}. Finally, a summary and concluding remarks are given in Sec.~\ref{sec:conclusion}.

%% file: Input.tex
\section{Analysis input}\label{sec:input}

%%%%%%%%%%%%%%%%%%%%%%%%%%%%%%%%%%%%%%%%%%%%%%%%%%%%%%%%%%%%%%%%%%%%%%
\subsection{Camera images}\label{sec:input:images}
\begin{figure}
  \center
  \includegraphics[width=5cm]{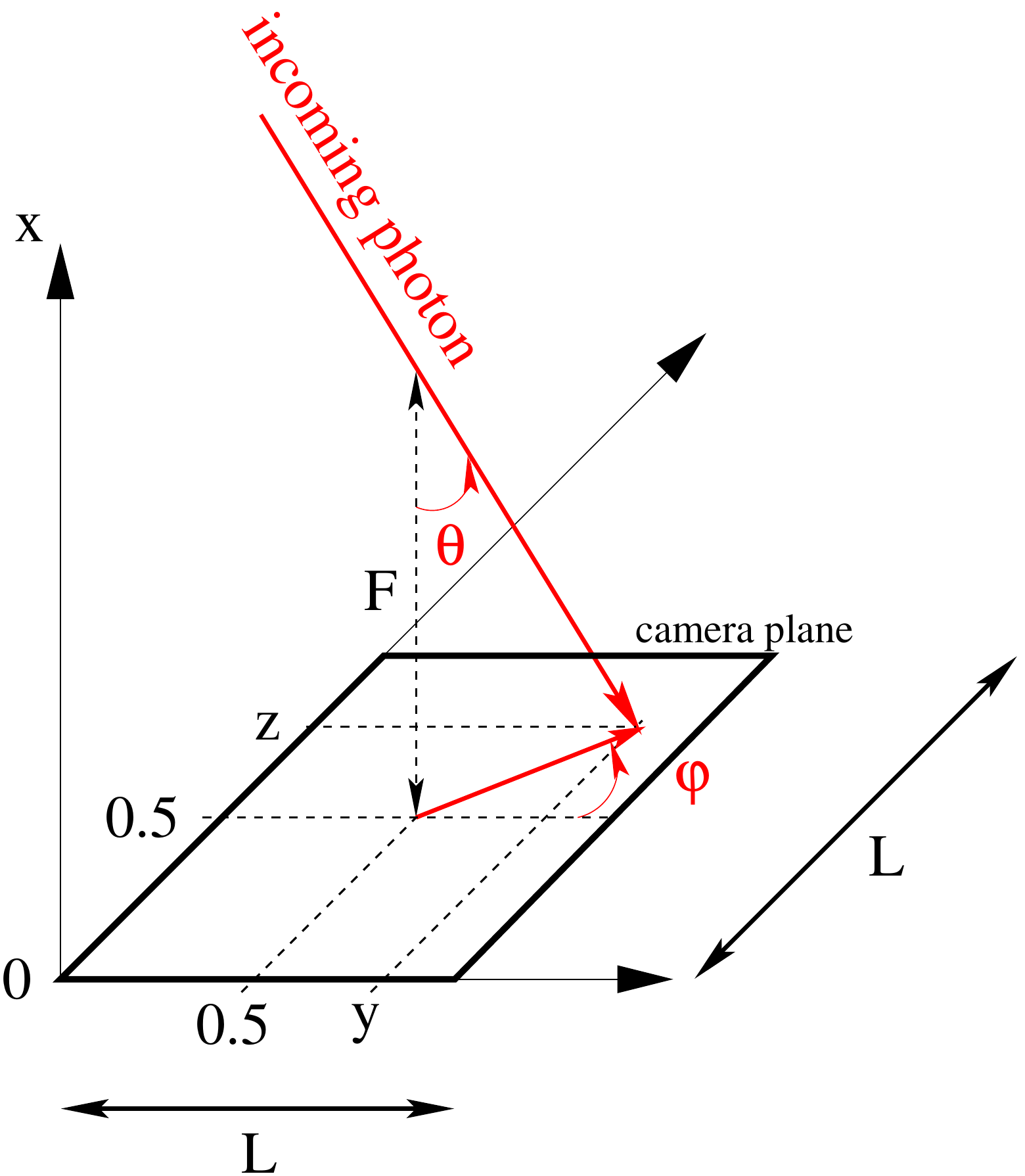}
  \includegraphics[width=8cm]{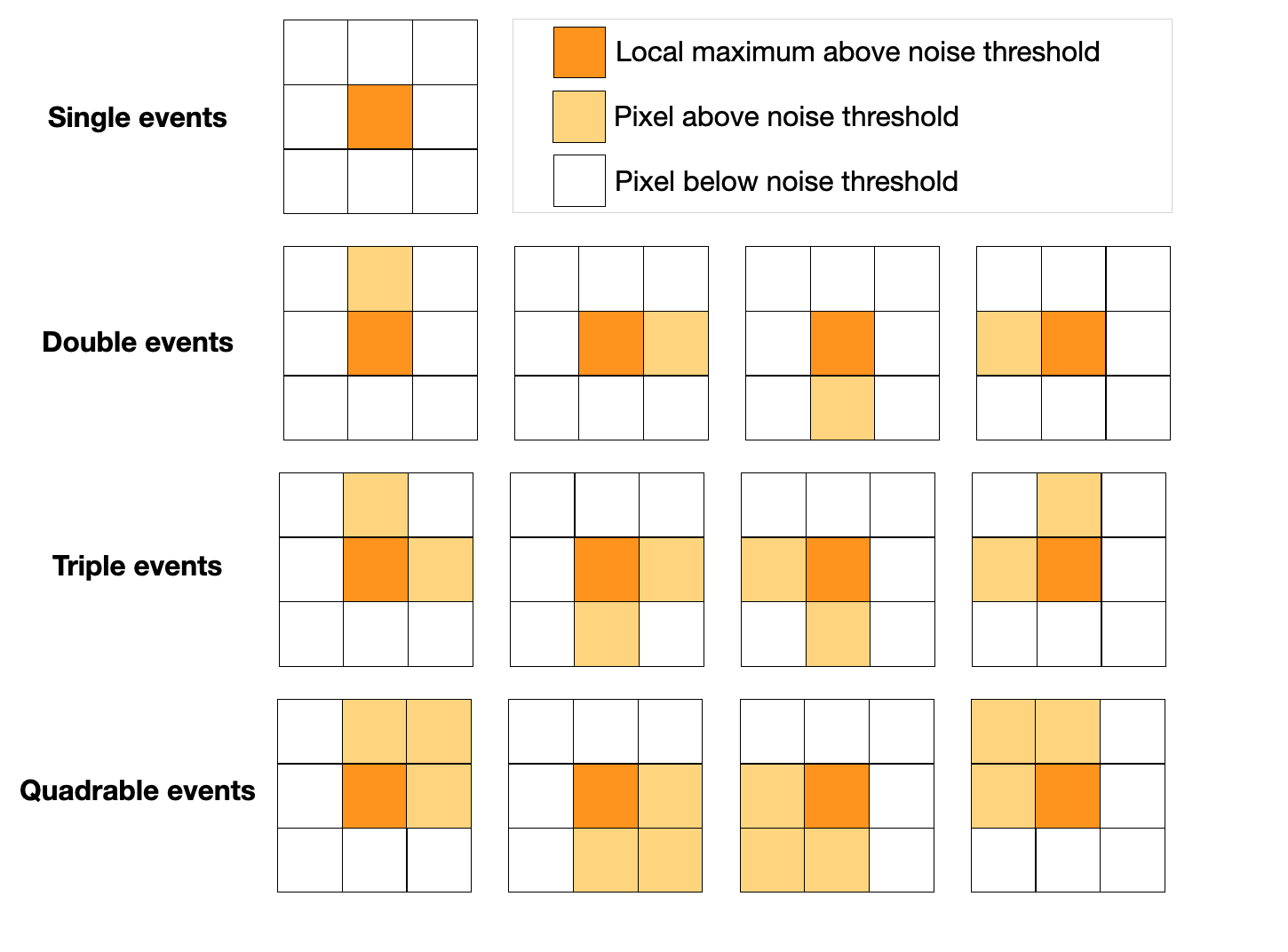}
  \caption{An incoming photon hits the camera plane (left). From the hit position $(y,z)$, the spherical angle $(\theta, \varphi)$ of the incoming direction can be derived. The left-hand diagram shows the most probable pixel patterns induced by a photon hitting the camera plane.}
  \label{fig:mxtdetctor}
\end{figure}

The MXT camera is a pn-Charge Coupled Device (pnCCD)~\citep{schneider2022,CERAUDO2020164164} installed in the telescope $yz$ focal plane. The $x$ axis is the optical axis of the telescope. This is represented schematically in the left-hand side of Fig.~\ref{fig:mxtdetctor}. The camera pixels are indexed by $i$ and $j$, running from $0$ to $N-1=255$ along the $y$ (resp. $z$) axis. We also use intrinsic coordinates, $0\le y <1$ and $0\le z <1$, to measure continuous positions in the camera plane.

An incoming X-ray photon interacts with the MXT camera pixels and deposits its energy via photo-electric effect. This energy is distributed over one or a few contiguous pixels, forming a pattern. The most probable pixel patterns induced by a photon and after thresholding on the pixel amplitude are represented in the right-hand side diagram in Fig.~\ref{fig:mxtdetctor}~\citep{Godet:2008ma}.

The MXT camera is operated in two modes. In \textit{full-frame} mode, the camera front-end electronics capture an image every 200~ms and transfer it to the MXT data processing unit. The full-frame images are processed to estimate the pixel noise and derive pixel-by-pixel detection thresholds. For astrophysical observations, the camera is set to \textit{event} mode. The image is integrated over 100~ms and the pixel noise component is removed by the front-end electronics~\citep{CERAUDO2020164164}. When the amplitude is above the detection threshold, the pixel data is transfered to the MXT data processing unit for analysis.

In event mode, the first analysis step is to cluster adjacent pixels together. A photon is identified if the pixel pattern matches one of those represented in the right-hand side of Fig. \ref{fig:mxtdetctor}. The other pixel patterns are attributed to noise (e.g. cosmic rays) and are discarded. Photons are parametrized by the energy $E_p=\sum_{ij}{E[i][j]}$ integrated over the $(i,j)$ pixels in the cluster. The photon hit position $(y_p, z_p)$ in intrinsic coordinates is estimated by computing the energy-weighted pixel position:
\begin{align}
  y_p &= \frac{\sum_{ij} E[i][j] \times (i + 0.5)}{ N \sum_{ij} E[i][j] }\quad \mathrm{and}\\
  z_p &= \frac{\sum_{ij} E[i][j] \times (j + 0.5)}{ N \sum_{ij} E[i][j] },
  \label{eq:photonposition}
\end{align}
where the sums run over the pixels $(i,j)$ in the photon cluster.

As images are recorded, photons are cumulated onto a $N_D\times N_D$ counting map $D$ called the ``photon map''. To limit the computational cost, the resolution is set to $N_D=N/2=128$. It is shown in Sec.~\ref{sec:characterization:gridres} that the localization accuracy is hardly impacted by this choice of lower resolution. The right-hand plot of Fig.~\ref{fig:psfandcummap} shows an example of a photon map.

%%%%%%%%%%%%%%%%%%%%%%%%%%%%%%%%%%%%%%%%%%%%%%%%%%%%%%%%%%%%%%%%%%%%%%
\subsection{Point spread function}\label{sec:input:psf}
\begin{figure}
  \center
  \includegraphics[width=8cm]{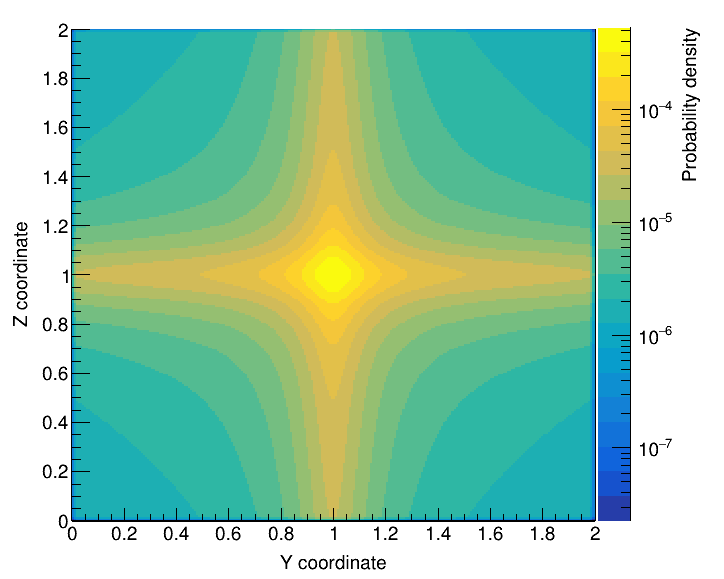}
  \includegraphics[width=8cm]{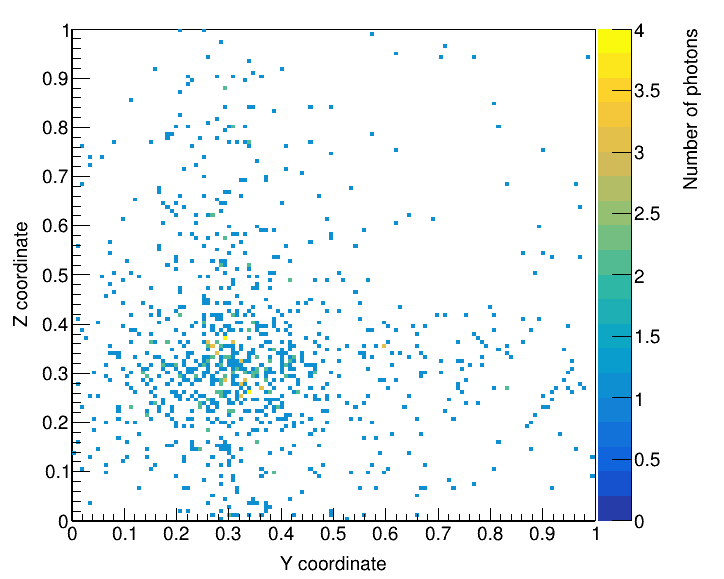}
  \caption{The MXT point spread function is represented on the left. An example of a resulting photon map $D$ with 1000 photons is represented on the right.
  }
  \label{fig:psfandcummap}
\end{figure}

The localization analysis presented in Sec.~\ref{sec:algorithm} takes the photon map $D$ as an input and cross-correlates it with the image of a point source produced by the MXT optical system. This expected image, called the point spread function, is measured and modeled on the ground and uploaded to the satellite. The point spread function of the MXT is the result of the microporous structure~\citep{10.1117/12.2561739} of its optical system. It measures the hit position probability of an incoming photon over the focal plane. It is composed of a central spot (double reflection), four cross arms (single reflection) and a diffused patch (no reflection). The MXT point spread function has been measured and modeled by the University of Leicester team, in charge of the MXT optical system. It is modeled by a function $P(y,z,E)$ fitted to X-ray data collected at different energies $E$ during the MXT performance tests conducted at the Panter facility, Neuried Germany, in November 2021. The flight-model point spread function used in this work, $P(y,z)$, is plotted in Fig.~\ref{fig:psfandcummap}. It is weighted over the expected number of photons for an average gamma-ray burst, taking into account a typical spectrum and the MXT photon collecting efficiency as a function of energy. The full width at half maximum of the central peak is about 11~arcmin ($\simeq 50$~pixels).

The point spread function $P$ is discretized over a $N_P\times N_P$ grid where $N_P=N=2N_D$. It is twice the size of the photon map in both directions to cover all possible incoming photon directions. Moreover, the point spread function is normalized to 1:
\begin{equation}
  \sum_{i=0}^{N_P-1}\sum_{j=0}^{N_P-1}{P[i][j]}=1.
  \label{eq:psfnorm}
\end{equation}
The right-hand side plot in Fig.~\ref{fig:psfandcummap} shows an example of a photon map generated with 1000 photons, placing the point spread function peak at $y=0.3$ and $z=0.3$. 

%% file: Algorithm.tex
\section{Localization algorithm}\label{sec:algorithm}
The localization analysis loops continuously in the MXT data processing unit. One cycle takes less than 2~s to complete. The analysis delivers the X-ray source sky location in spherical coordinates $(\theta,\varphi)$, as well as a photon count, a signal-to-noise ratio and a localization uncertainty. This information is updated every $~30$~s and packaged in telemetry packets sent to the ground. The photon map $D$, introduced in Sec.~\ref{sec:input:images}, is cross-correlated with the point spread function presented in Sec.~\ref{sec:input:psf}. As opposed to a simplistic barycenter/centroid approach, this method, described in Sec.~\ref{sec:algorithm:xcorr}, was selected as it offers a better sensitivity to faint sources. Indeed, one can take advantage of the specific MXT's point spread function for which the cross-arm pattern can guide the search for source peak positions.

The cross-correlation data is then processed to search for local maxima, as detailed in Sec.~\ref{sec:algorithm:position}. In Sec.~\ref{sec:algorithm:snr}, we explain how to estimate the source photon count and the background contribution. Finally, the algorithm is designed to locate the three brightest sources in the MXT field of view. To achieve this, an iterative subtraction process is implemented and is described in Sec.~\ref{sec:algorithm:multiple}.

%%%%%%%%%%%%%%%%%%%%%%%%%%%%%%%%%%%%%%%%%%%%%%%%%%%%%%%%%%%%%%%%%%%%%%
\subsection{Cross-correlation}\label{sec:algorithm:xcorr}
To determine the localization of the source, the photon map $D$ is cross-correlated with the point spread function $P$ such as:
\begin{equation}
  C[i][j] = \sum_{i^{\prime}=0}^{N_P-1} \sum_{j^{\prime}=0}^{N_P-1} D[i+i^{\prime}][j+j^{\prime}]\times P[i^{\prime}][j^{\prime}],
  \label{eq:xcorr1}
\end{equation}
where $i$ and $j$ take values between 0 and $N_D-1$. In Eq.~\ref{eq:xcorr1} the photon map is shifted by $(i,j)$ bins over the point spread function matrix and the cross-correlation coefficients are obtained by summing the products of all overlapping bin values in a $N_P^2$ grid. It is worth noting that the photon map indices may be out-of-range. As we are dealing with finite discrete functions, we assume circular periodicity where $i+i^\prime$ (resp. $j+j^\prime$) actually means $(i+i^\prime) \pmod {N_D}$ (resp. $(j+j^\prime) \pmod {N_D}$). In the following the out-of-range issues will no longer be addressed, and we will simply write $i+i^\prime$ (resp. $j+j^\prime$). As a result, in the sum of Eq.~\ref{eq:xcorr1} the photon map $D$ is used four times.

The implementation of the cross-correlation given in Eq.~\ref{eq:xcorr1} is computationally expensive. Alternatively, the cross-correlation can be computed in the Fourier space:
\begin{equation}
    \tilde{C}[k][l] = \frac{1}{N_D^2}\sum_{i=0}^{(N_D-1)}\sum_{j=0}^{(N_D-1)}C[i][j]\times e^{-2\sqrt{-1}\pi(ik+jl)/N_D}.
    \label{eq:xcorr:fourier0}
\end{equation}
According to the correlation theorem \citep{papoulis62}, the cross-correlation of two signals is equivalent to a complex conjugate multiplication of their Fourier transforms:
\begin{equation}
\tilde{C}[k][l] = N_P^2 \times \tilde{D}[k][l] \times \tilde{P}^*[2k][2l],
  \label{eq:xcorr:fourier1}
\end{equation}
where we use $^*$ for the complex conjugate. In Eq.~\ref{eq:xcorr:fourier1}, both the photon map $\tilde{D}$ and the cross-correlation matrix $\tilde{C}$ are sampled at a frequency $1/N_D$, while the point spread function matrix is sampled at a frequency $1/N_P=1/(2N_D)$:
\begin{equation}
    \tilde{P}[k][l] = \frac{1}{N_P^2}\sum_{i=0}^{(N_P-1)}\sum_{j=0}^{(N_P-1)}P[i][j]\times e^{-2\sqrt{-1}\pi(ik+jl)/N_P}.
    \label{eq:psf:fourier}
\end{equation}
As a consequence, every other coefficients of $\tilde{P}$ are considered in Eq.~\ref{eq:xcorr:fourier1}. 

Because of the limited computing power on-board and to optimize the uplink bandwidth, the point spread function is formatted on the ground and the final $N_P^2\times\tilde{P}^*[2k][2l]$ table is uploaded onboard with a telecommand. Having one Fourier coefficient over two, the on-board point spread function is therefore partial. This limitation has important consequences to compute the signal and background counts: this is discussed in Sec.~\ref{sec:algorithm:snr}. The cross-correlation map is obtained by applying an inverse Fourier Transform to Eq.~\ref{eq:xcorr:fourier1}:
\begin{equation}
  C[i][j] = N^2_P\times \sum_{k=0}^{(N_D-1)}\sum_{l=0}^{(N_D-1)} \tilde{D}[k][l] \times\tilde{P}^*[2k][2l]\times e^{+2\sqrt{-1}\pi(ik+jl)/N_D}.
  \label{eq:xcorr:fourier3}
\end{equation}

 This cross-correlation method has two systematic biases which must be corrected. The first bias is introduced when working with a point spread function twice the size of the photon map. The photon map must be shifted by half of a bin in both directions to be aligned with the point spread function. Therefore the cross-correlation map is also shifted by half of a bin. When localizing a peak in the cross-correlation map, an offset of $0.5/N_D$ must be removed in $y$ and $z$. The second bias is caused by the Fourier transform spectral leakage: it is more pronounced when the source peak stands near the edges of the camera plane. This bias is discussed and corrected in Sec.~\ref{sec:characterization:edges}

%%%%%%%%%%%%%%%%%%%%%%%%%%%%%%%%%%%%%%%%%%%%%%%%%%%%%%%%%%%%%%%%%%%%%%
\subsection{Source angular position}\label{sec:algorithm:position}
The X-ray source position is associated to a peak in the cross-correlation map $C$. First, the global maximum $C[i_1][j_1]$ in the cross-correlation map is identified. Then we define a window centered on this maximum: $i_1-N_w\le i \le i_1+N_w$ and $j_1-N_w\le j \le j_1+N_w$. As explained in Sec.~\ref{sec:algorithm:xcorr}, we use circular indices if the window overlaps the edges of the cross-correlation map. The size of the window must be chosen to fully include the central spot of the point spread function. For the MXT point spread function, we use $N_w=25$. The peak position $(y_1,z_1)$ is finally computed as a two-dimensional barycenter inside the window:
\begin{equation}
  y_1 = \frac{\sum_{i,j} C[i][j] \times (i+0.5)}{N_D\times\sum_{i,j} C[i][j]},\quad \mathrm{and} \quad
  z_1 = \frac{\sum_{i,j} C[i][j] \times (j+0.5)}{N_D\times\sum_{i,j} C[i,j]}.
  \label{eq:peakposition}
\end{equation}
The sums run over indices $(i,j)$ inside the window and where the cross-correlation coefficients take significant values: $C[i][j] > \alpha \times C[i_1][j_1]$. Using simulated data, we find that $\alpha = 0.9$ gives the best localization accuracy. It is worth noting that both the $\alpha$ and $N_w$ parameters are configurable from the ground.

\begin{figure}
  \center
  \includegraphics[width=7cm]{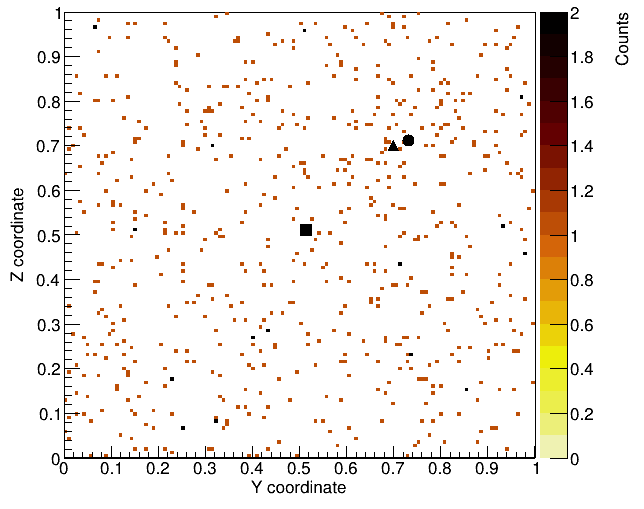}
  \includegraphics[width=7cm]{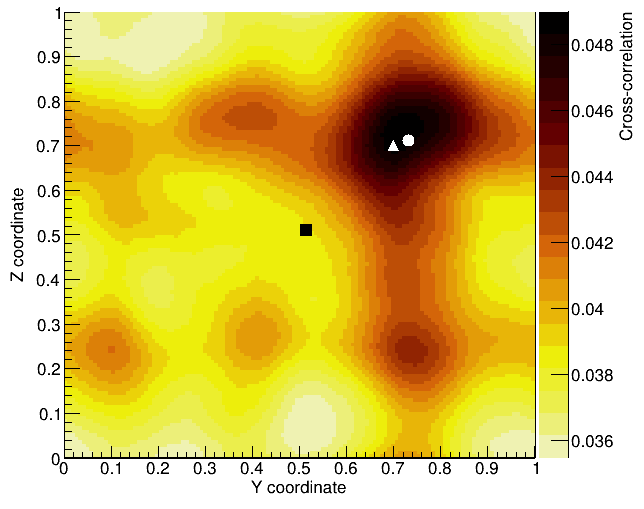}
  \caption{Example of a photon map $D$ simulated with a flat background of 600 counts and a faint source positioned at $y=0.7$ and $z=0.7$ (triangle marker) with 50 counts (right plot). The source position is well visible in the cross-correlation map (right). It is localized at $y=0.73$ and $z=0.72$ by the MXT on-board analysis (circular marker). A simple barycenter fails at localizing the source (square marker).}
  \label{fig:corrAndcum}
\end{figure}

Figure~\ref{fig:corrAndcum} shows an example of a photon map simulated with a faint source (50 photons) on top of a uniform background (600 counts). The source position cannot be identified in the photon map while it appears clearly in the cross-correlation map. The peak position is localized reasonably well with the method described above (circular marker). The true position is indicated with a triangular marker. For comparison, we also add the position derived from a simple barycenter evaluated over the entire photon map (square marker) which fails at localizing the source.

As introduced in Sec.~\ref{sec:algorithm:xcorr}, the cross-correlation map suffers from two systematical biases: the Fourier transform spectral leakage and the alignment asymmetry between the photon map and the point spread function. For the first effect we apply \textit{ad hoc} corrections $\eta_1(y)$ and $\eta_2(z)$ which are evaluated and discussed in Sec.~\ref{sec:characterization:edges}. The second bias is corrected by applying a shift of half of a bin:
\begin{equation}
  y^{\prime}_1 = y_{1} + \eta_1(y_1) - \frac{0.5}{N_D}, \quad \mathrm{and} \quad
  z^{\prime}_1 = z_{1} + \eta_2(z_1) - \frac{0.5}{N_D}.
  \label{eq:peakposition_corrected}
\end{equation}

Finally, the source position ($y^{\prime}_1, z^{\prime}_1$) is converted to spherical angles $(\theta, \varphi)$ using the telescope focal length $F$ and the physical size of the camera $L\times L$:
\begin{align}
  \theta &= \tan^{-1}{ \left(\frac{L}{F} \times \sqrt{(y^{\prime}_{1} - 0.5)^2 + (z^{\prime}_1 - 0.5)^2}\right)}, \nonumber \\
  \varphi &= \tan^{-1}\left(\frac{z^{\prime}_1-0.5}{y^{\prime}_1- 0.5}\right).
  \label{eq:thetaphi}
\end{align}

%%%%%%%%%%%%%%%%%%%%%%%%%%%%%%%%%%%%%%%%%%%%%%%%%%%%%%%%%%%%%%%%%%%%%%
\subsection{Signal and noise counts}\label{sec:algorithm:snr}
The source signal $S_1$ and noise $B$ counts are estimated on board the MXT to derive the source signal-to-noise ratio $\rho_1 = S_1/\sqrt{B}$. A standard method to estimate these quantities consists of analyzing the photon map. The signal+noise component is integrated around the main peak and the noise-only component is estimated using a region of the image where the signal contribution can be neglected. For the MXT, this method is not optimal due to the cross-shaped structure of the point spread function and the possible presence of multiple X-ray sources in the field of view. Moreover there are large uncertainties associated to faint signals and low-statistic backgrounds.

\begin{figure}
  \center
  \includegraphics[width=6cm]{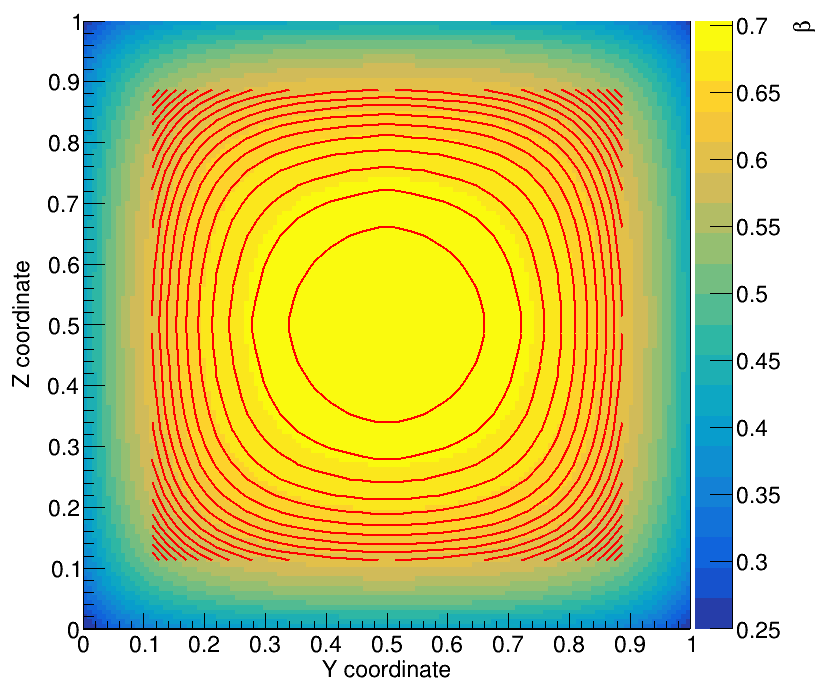}
  \includegraphics[width=6cm]{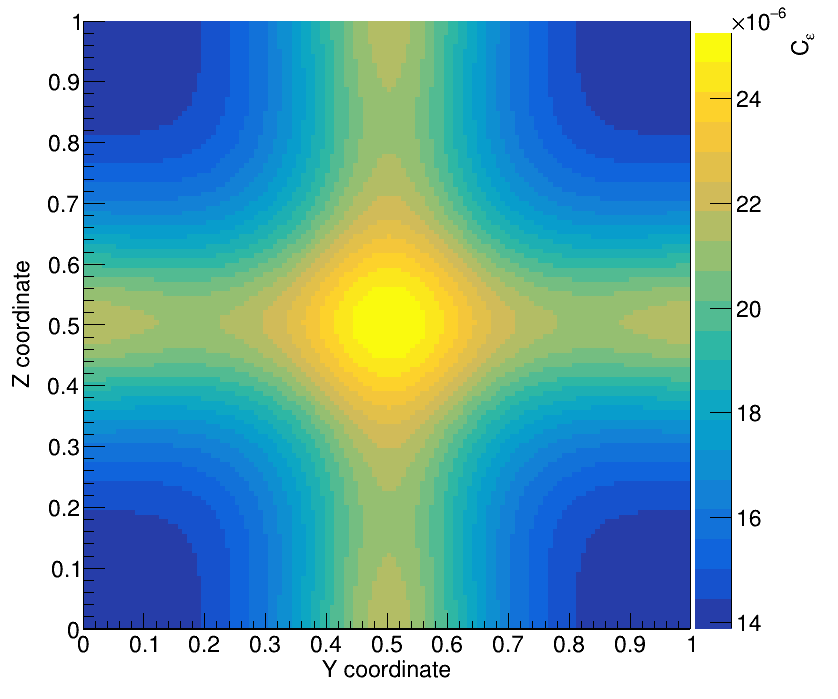}
  \caption{For a given point spread function, the $\beta$ function (left) is measured using Eq.~\ref{eq:beta}. It is then fitted: the fit is represented by the red contour. The function $C_{\varepsilon}$ (right) is measured with $B=0$ (Eq.~\ref{eq:c1}) and a source positioned at the center.}
  \label{fig:beta_cepsilon}
\end{figure}

Instead, we use the cross-correlation map $C$ which naturally integrates the background and signal counts over the entire camera plane. Decomposing the photon map $D$ in a linear combination of a signal component $s_1$ and a background component, we can write Eq.~\ref{eq:xcorr1} as:
\begin{align}
  C[i][j] &= \sum_{i^{\prime}=0}^{N_P-1} \sum_{j^{\prime}=0}^{N_P-1} \left(s_1[i^{\prime}][j^{\prime}]\times P[i^{\prime}+i][j^{\prime}+j]\right) + \frac{B_1}{N_D^2},
  \label{eq:xcorr_1}
\end{align}
where we assume the background $B_1$ to be uniformly distribued over the camera plane and where we used the normalization of Eq.~\ref{eq:psfnorm}. The source distribution is driven by the point spread function:
\begin{align}
  s_1[i][j] &= \frac{S_1}{\sum_{i^\prime=0}^{N_D-1}\sum_{j^\prime=0}^{N_D-1}P[N_D+i^\prime-i_1][N_D+j^\prime-j_1]}\times P[N_D+i-i_1][N_D+j-j_1], \nonumber \\
  &= \frac{S_1}{\beta[i_1][j_1]}\times P[N_D+i-i_1][N_D+j-j_1],
  \label{eq:s1}
\end{align}
where $i_1=\lfloor y_1^{\prime}N_D \rfloor$ and $j_1=\lfloor z_1^{\prime}N_D \rfloor$ are the indices matching the cross-correlation peak position $(y_1^{\prime},z_1^{\prime})$ obtained in Eq.~\ref{eq:peakposition_corrected} and where we introduce:
\begin{equation}
\beta[i_1][j_1] = \sum_{i^\prime=0}^{N_D-1}\sum_{j^\prime=0}^{N_D-1}P[N_D+i^\prime-i_1][N_D+j^\prime-j_1].
  \label{eq:beta}
\end{equation}
The $\beta$ function cannot be computed exactly on board as the point spread function is incomplete; every other Fourier coefficients $\tilde{P}^*[2k][2l]$ are uploaded. Instead, the $\beta$ function is approximated by a fit function, the parameters of which are uploaded: see the left plot in Fig.~\ref{fig:beta_cepsilon}.

Similarly, the signal component cannot be calculated completely so we re-write Eq~\ref{eq:s1} as:
\begin{equation}
  s_1[i][j] = \frac{S_1}{\beta[i_1][j_1]}\times \left(P_{\mathrm{comp}}[i-i_1][j-j_1] - \varepsilon_1[i][j]\right),
\end{equation}
where we introduce a ``compact'' point spread function of size $N_D\times N_D$:
\begin{equation}
  P_{\mathrm{comp}}[i][j] = P[i][j] + P[i][N_D+j] + P[N_D+i][j] + P[N_D+i][N_D+j],
  \label{eq:psfcomp}
\end{equation}
the Fourier transform of which is:
\begin{equation}
  \tilde{P}_{\mathrm{comp}}[k][l] = 4\times\tilde{P}[2k][2l].
  \label{eq:psfcomptilde}
\end{equation}
The $\varepsilon_1$ function describes the missing terms of the point spread function in Eq.~\ref{eq:s1}. The cross-correlation of the signal component with the point spread function is:
\begin{align}
  C_1[i][j] &= \frac{S_1}{\beta[i_1][j_1]}\times\sum_{i^{\prime}=0}^{N_P-1} \sum_{j^{\prime}=0}^{N_P-1} P[i^{\prime}+i][j^{\prime}+j]\times \left(P_{\mathrm{comp}}[i^{\prime}-i_1][j^{\prime}-j_1] - \varepsilon_1[i^{\prime}][j^{\prime}]\right) \nonumber \\
  &= \frac{S_1}{\beta[i_1][j_1]}\times (W[i-i_1][j-j_1] - C_{\varepsilon_1}[i][j]).
  \label{eq:c1}
\end{align}
This cross-correlation coefficient has two components. The first one, $W$, is the cross-correlation of $P_{\mathrm{comp}}$ with the point spread function. The $W$ function is computed on-board substituting $D$ by $P_{\mathrm{comp}}$ in Eq.~\ref{eq:xcorr:fourier3}:
\begin{equation}
  W[i][j] = 4N^2_P\times \sum_{k=0}^{(N_D-1)}\sum_{l=0}^{(N_D-1)} \tilde{P}[2k][2l] \times\tilde{P}^*[2k][2l]\times e^{+2\sqrt{-1}\pi(ik+jl)/N_D}.
  \label{eq:autocorr}
\end{equation}
The second component, $C_{\varepsilon_1}$, is the cross correlation between $\varepsilon_1$ and the point spread function. It is unknown.

Let us consider two points in the cross-correlation map $C$: the cross-correlation peak at $(i_1, j_1)$ and the cross-correlation minimum point at $(i_m, j_m)$. We get:
\begin{equation}
  \begin{cases}
    C[i_1][i_1] = \frac{S_1}{\beta[i_1][j_1]}\times (W[0][0] - C_{\varepsilon_1}[i_1][j_1]) + \frac{B_1}{N_D^2} \\
    C[i_m][i_m] = \frac{S_1}{\beta[i_1][j_1]}\times (W[i_m-i_1][j_m-j_1] - C_{\varepsilon_1}[i_m][j_m]) + \frac{B_1}{N_D^2}
  \end{cases}
\end{equation}
Solving this system of equations we get:
\begin{align}
  S_1 &= \beta[i_1][j_1]\times\frac{C[i_1][j_1]-C[i_m][j_m]}{W[0][0] - C_{\varepsilon_1}[i_1][j_1] - W[i_m-i_1][j_m-j_1] + C_{\varepsilon_1}[i_m][j_m]}, \label{eq:s1_estimate}\\
  B_1 &= N_D^2\times \left(C[i_1][j_1]-\frac{S_1}{\beta[i_1][j_1]}\times (W[0][0]-C_{\varepsilon_1}[i_1][j_1])\right). \label{eq:b_estimate}
\end{align}
To estimate $S_1$ and $B_1$ with Eqs.~\ref{eq:s1_estimate} and~\ref{eq:b_estimate}, we approximate the $C_{\varepsilon_1}$ values by:
\begin{align}
  C_{\varepsilon_1}[i_m][j_m] - C_{\varepsilon_1}[i_1][j_1]&= \mathrm{Min}_{i,j}(C_{\varepsilon}[i][j]) - \mathrm{Max}_{i,j}(C_{\varepsilon}[i][j]) \quad \mathrm{and} \label{eq:cepsilon_s}\\
  C_{\varepsilon_1}[i_1][j_1] &= \mathrm{Max}_{i,j}(C_{\varepsilon}[i][j]) \label{eq:cepsilon_b},
\end{align}
where $C_{\varepsilon}$ is a cross-correlation function computed on the ground with a source positioned at the center of the camera plane: see the right plot in Fig.~\ref{fig:beta_cepsilon}.

%%%%%%%%%%%%%%%%%%%%%%%%%%%%%%%%%%%%%%%%%%%%%%%%%%%%%%%%%%%%%%%%%%%%%%
\subsection{Localization of multiple sources}\label{sec:algorithm:multiple}
Given the detector sensitivity~\citep{MXTGotz:2015mha}, it is possible to have more than one X-ray source within the field of view of the MXT. Using data collected by the ROSAT telescope~\citep{Boller:2016pne} which was operated in an energy band comparable to the MXT and had a similar sensitivity, we expect to have at most three X-ray sources within the field of view of the MXT. As a result, the localization algorithm shall be able to manage multiple sources. The localization analysis described in Secs.~\ref{sec:algorithm:xcorr},~\ref{sec:algorithm:position}, and~\ref{sec:algorithm:snr} is repeated three times, subtracting the contribution of the detected source at each iteration.

The cross-correlation map at iteration $g=1$ is $C^{g=1} = C$, where $C$ is computed with Eq.~\ref{eq:xcorr:fourier3}. For the next two iterations, $g=2$ and $g=3$, we use $C^g=C^{g-1}-C_{g-1}$, where the contribution of source $g$, $C_g$, is derived from Eq.~\ref{eq:c1}. Note that the $C_{\varepsilon_g}$ function is approximated by a single value: $C_{\varepsilon_g}[i][j]\simeq \mathrm{Max}_{i,j}(C_{\varepsilon}[i][j])$.

After three iterations, we have three estimates for the background: $B_1$, $B_2$, and $B_3$. None of them is a perfect estimate of $B$. For a single source, the best estimator is $B_1$. For multiple sources, however, the choice is not trivial: the first estimate $B_1$ is over-estimated by secondary sources, while the last estimate $B_3$ is under-estimated by imperfect subtractions. As a compromise, the background is updated at each iteration:
\begin{equation}
  B = B_g\quad \mathrm{if}\quad S_g > \rho_q\times \sqrt{B_g},
\end{equation}
where $\rho_q$ is a configurable parameter.

\begin{figure}
  \center
  \includegraphics[width=6cm]{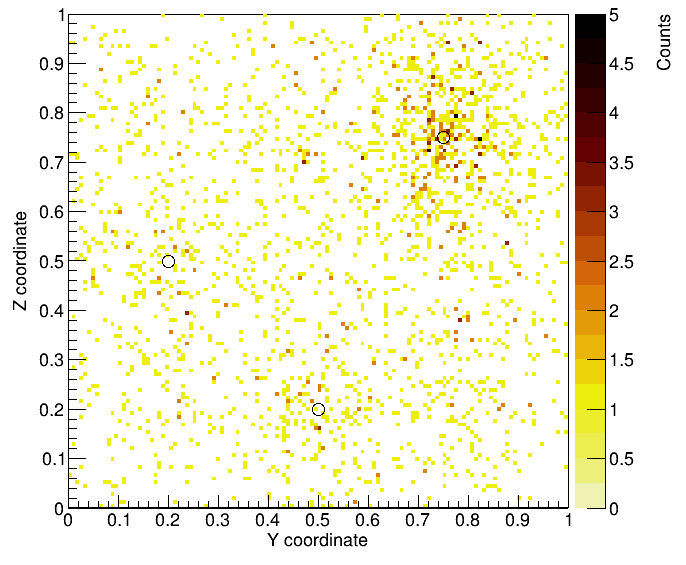}
  \includegraphics[width=6cm]{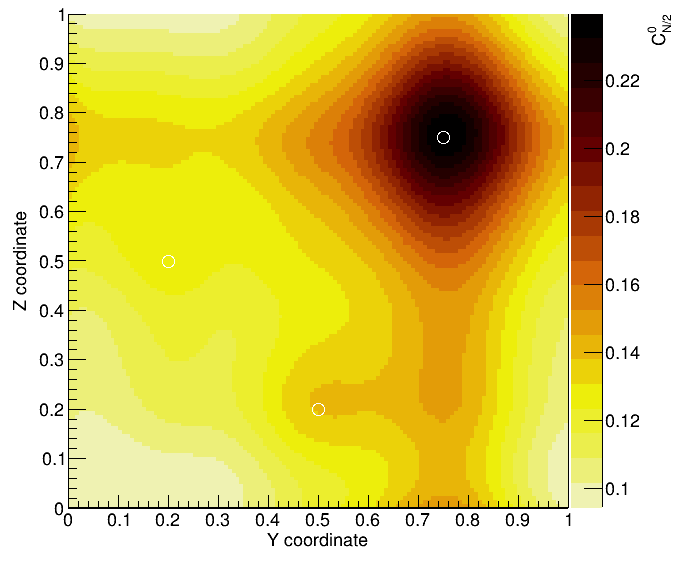} \\
  \includegraphics[width=6cm]{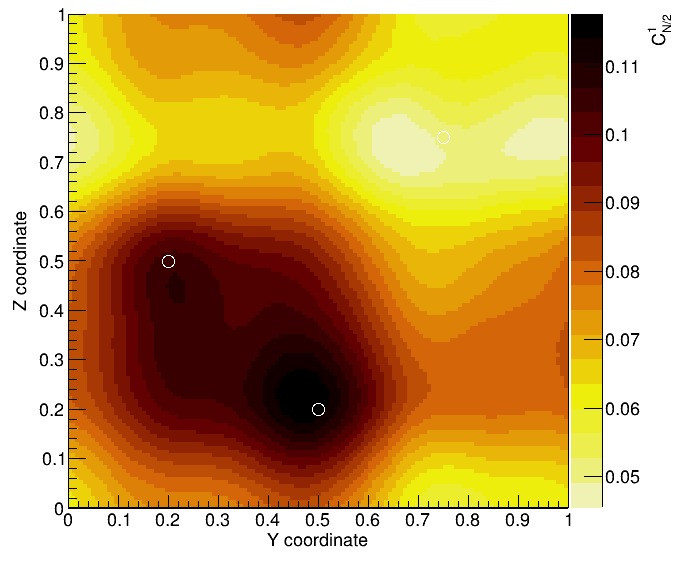}
  \includegraphics[width=6cm]{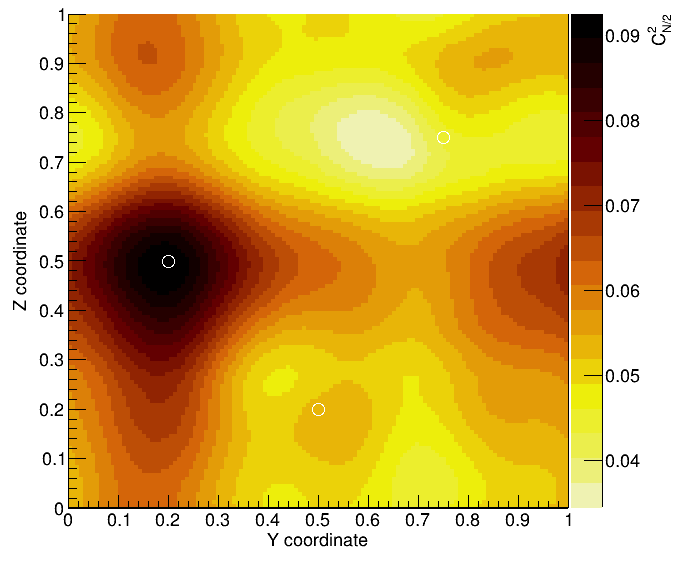}
  \caption{The top-left plot shows the distribution of photons from three separate sources ($S_1=1000$, $S_2=300$, and $S_3=300$) on top of a uniform background ($B=600$). The three other plots show the initial cross-correlation map $C^1$ (top-right), after subtracting the first source $C^2$ (bottom-left), and after subtracting the second source $C^3$ (bottom-right). The true source positions are indicated by the circles.}
  \label{fig:multiple}
\end{figure}

Figure~\ref{fig:multiple} illustrates the subtraction method. The photon map $D$ includes three photon sources simulated with $S_1=1000$, $S_2=300$, and $S_3=300$. The background is uniform and set to $B=600$. Running our analysis, these counts are estimated by $S_1=1045$ (+4.5\%), $S_2=386$ (+29\%), $S_3=299$ (+0.0\%), $B_1=824$, $B_2=810$, and $B_3=627$. The final background is estimated at $B=B_3=627$ (+4.5\%).

%% file: Characterization.tex
\section{Characterization of the localization method}\label{sec:characterization}
The localization algorithm presented in Sec.~\ref{sec:algorithm} is characterized using different test scenarios. For each scenario we generate camera images with the \texttt{SatAndLight} simulation toolkit~\citep{satandlight}, which is designed to include many effects: photons from astrophysical and background sources, cosmic rays, corrupted camera pixels, etc. Astrophysical sources generate photons at a given energy, or following a given energy distribution. Photons are reflected by the MXT optical system following the point spread function (see Sec.~\ref{sec:input:psf}) and interact with the camera CCD pixels. A cosmic X-ray background is simulated by injecting high-energy photons uniformly on the camera plane. For the MXT, we expect to detect one background photon every second. In the following we will often consider a canonical background of $B=600$ counts, corresponding to an observation of 10 minutes.

%%%%%%%%%%%%%%%%%%%%%%%%%%%%%%%%%%%%%%%%%%%%%%%%%%%%%%%%%%%%%%%%%%%%%%
\subsection{The source intensity}\label{sec:characterization:r90}
In this section, the localization accuracy and its associated uncertainty is evaluated for different source and noise intensities.
We use the $r_{90}$ quantity,
\begin{equation}
  r_{90} = \left[\angle(\vec{r}_{\mathrm{meas}}, \vec{r}_{\mathrm{true}}) \right]_{90\%},
\end{equation}
which measures the angular distance between the source position recontructed by the localization algorithm $\vec{r}_{\mathrm{meas}}$ (Eq.~\ref{eq:peakposition_corrected}) and the true source position defined in the simulation $\vec{r}_{\mathrm{true}}$. This angular distance is measured with 1000 simulations of different positions $\vec{r}_{\mathrm{true}}$ randomly drawn in the field of view of the telescope. The $r_{90}$ quantity is obtained from the 90$^{\mathrm{th}}$ percentile of the distribution of $\angle(\vec{r}_{\mathrm{meas}}, \vec{r}_{\mathrm{true}})$.

\begin{figure}
  \center
  \includegraphics[width=12cm]{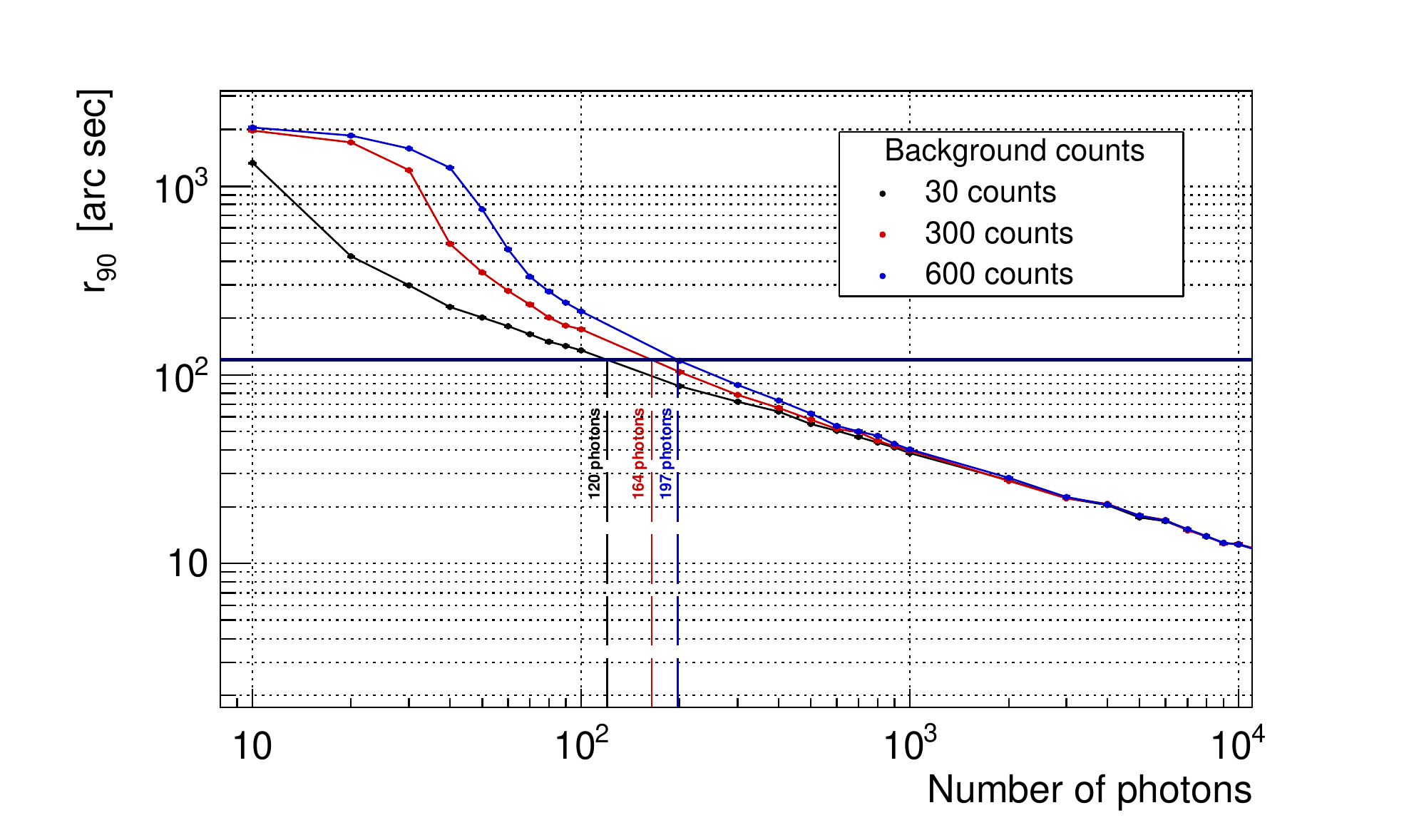}
  \caption{Localization uncertainty $r_{90}$ evaluated for a source intensity ranging from 10 to 10000 photons and with different background levels. Each point in this graph results from 1000 source positions randomly drawn in a $17\times 17$~arcmin$^2$ region around the camera center. Here, we use $N_D=N=256$.}
  \label{fig:r90_256}
\end{figure}

The MXT was designed to achieve a source localization accuracy of $r_{90} < 2$~arcmin for 85\% of the detected gamma-ray bursts. In Fig.~\ref{fig:r90_256}, $r_{90}$ is evaluated for an X-ray source with an intensity ranging from 10 to 10000 photons. The source photons are cumulated on top of a background noise of different intensities: 30, 300, and 600 counts. For these three background levels, the localization uncertainty is below 2~arcmin for 120, 164, and 197 source photons respectively.

%%%%%%%%%%%%%%%%%%%%%%%%%%%%%%%%%%%%%%%%%%%%%%%%%%%%%%%%%%%%%%%%%%%%%%
\subsection{The photon map resolution}\label{sec:characterization:gridres}
After photons are reconstructed, as explained in Sec.~\ref{sec:input:images}, they are cumulated onto a $N_D\times N_D$ two-dimensional grid called the photon map. The resolution of the photon map must be kept at minimum to limit the computing cost associated to the Fourier transform used to cross-correlate the data with the point spread function (see Sec.~\ref{sec:algorithm:xcorr}). However, it should be chosen high enough to fully resolve the peak of the point spread function and maximize the localization performance.

\begin{figure}
  \center
  \includegraphics[width=12cm]{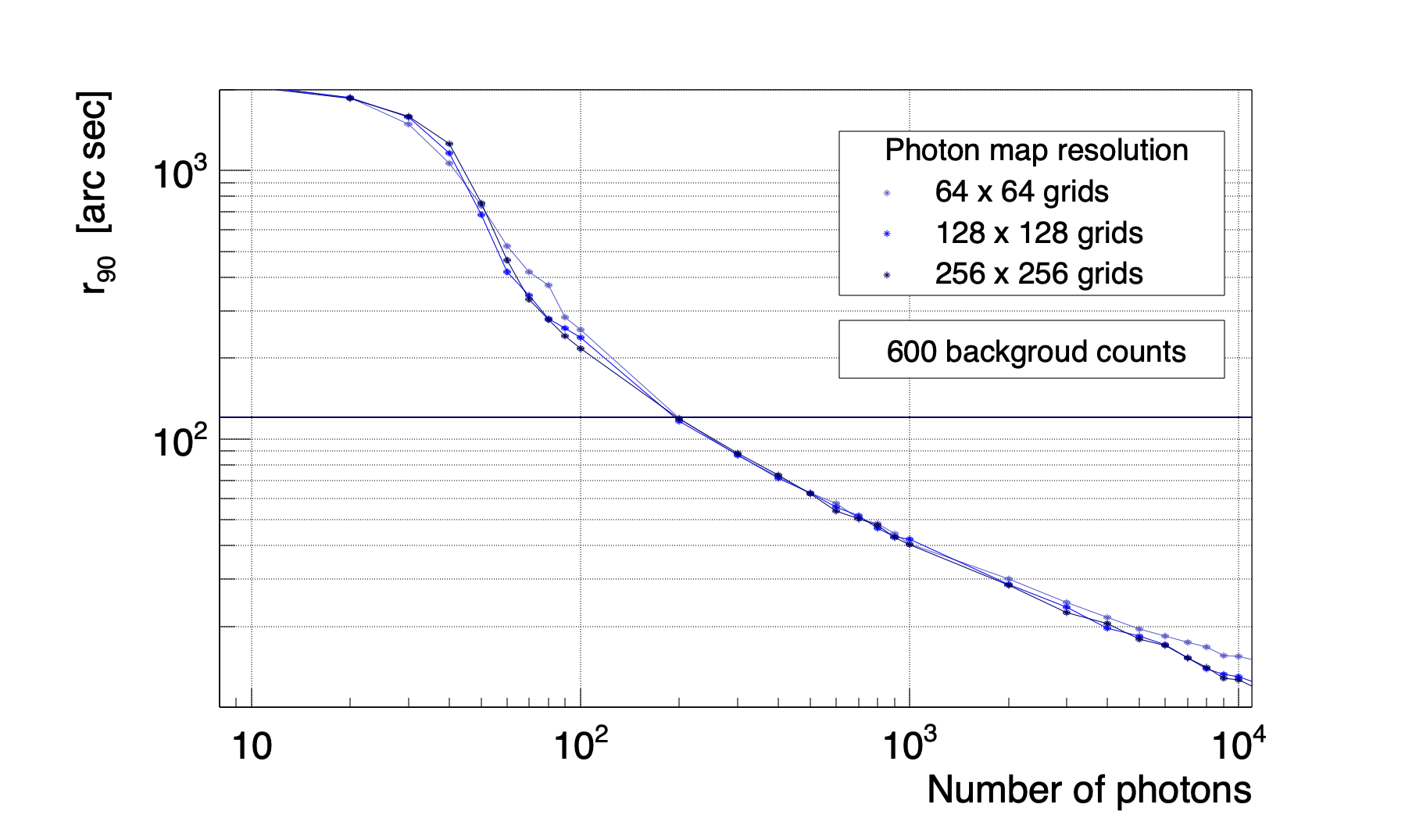}
  \caption{Localization uncertainty $r_{90}$ vs. number of photons evaluated for various photon map resolutions: $64\times 64$ (light blue), $128\times 128$ (bright blue), and $256\times 256$ (dark blue). The background level is fixed to 600 counts.}
  \label{fig:r90_res}
\end{figure}

To drive this choice, we measure the impact of the photon map resolution on the localization performance. The results are presented in Fig.~\ref{fig:r90_res}. There is no significant impact between the full resolution ($N_D=N=256$) and using a $128\times 128$ grid. We observe a $\sim 8\%$ loss when the resolution is further reduced ($64\times 64$). We only consider power-of-two values for $N_D$ to perform optimal Fourier transforms. For the MXT localization algorithm, and for the rest of this paper, we fix $N_D=128$. This resolution is sufficiently high to completely sample the shape of the point spread function and to offer an optimal localization accuracy. To perform the cross-correlation, we developed a two-dimensional $128\times 128$ complex-to-complex Fourier transform algorithm. It takes approximately 300~ms to complete one Fourier transform with the on-board computer.

%%%%%%%%%%%%%%%%%%%%%%%%%%%%%%%%%%%%%%%%%%%%%%%%%%%%%%%%%%%%%%%%%%%%%%
\subsection{Detector edge bias}\label{sec:characterization:edges}
The cross-correlation between the photon map and the point spread function is performed in the Fourier domain (Eq.~\ref{eq:xcorr:fourier1}). The discrete Fourier transform of size $N_D$ assumes periodicity: $D[0]\simeq D[N_D-1]$ in both the $y$ and $z$ directions. This assumption is reasonable when the X-ray source is positioned around the center of the field of view. It is highly violated when the source peak gets closer to the camera edges, when a fraction of the peak is chopped off. In this situation, spectral-leakage effects can be large and they bias the peak position in the cross-correlation map $C$.

\begin{figure}
  \center
  \includegraphics[width=12cm]{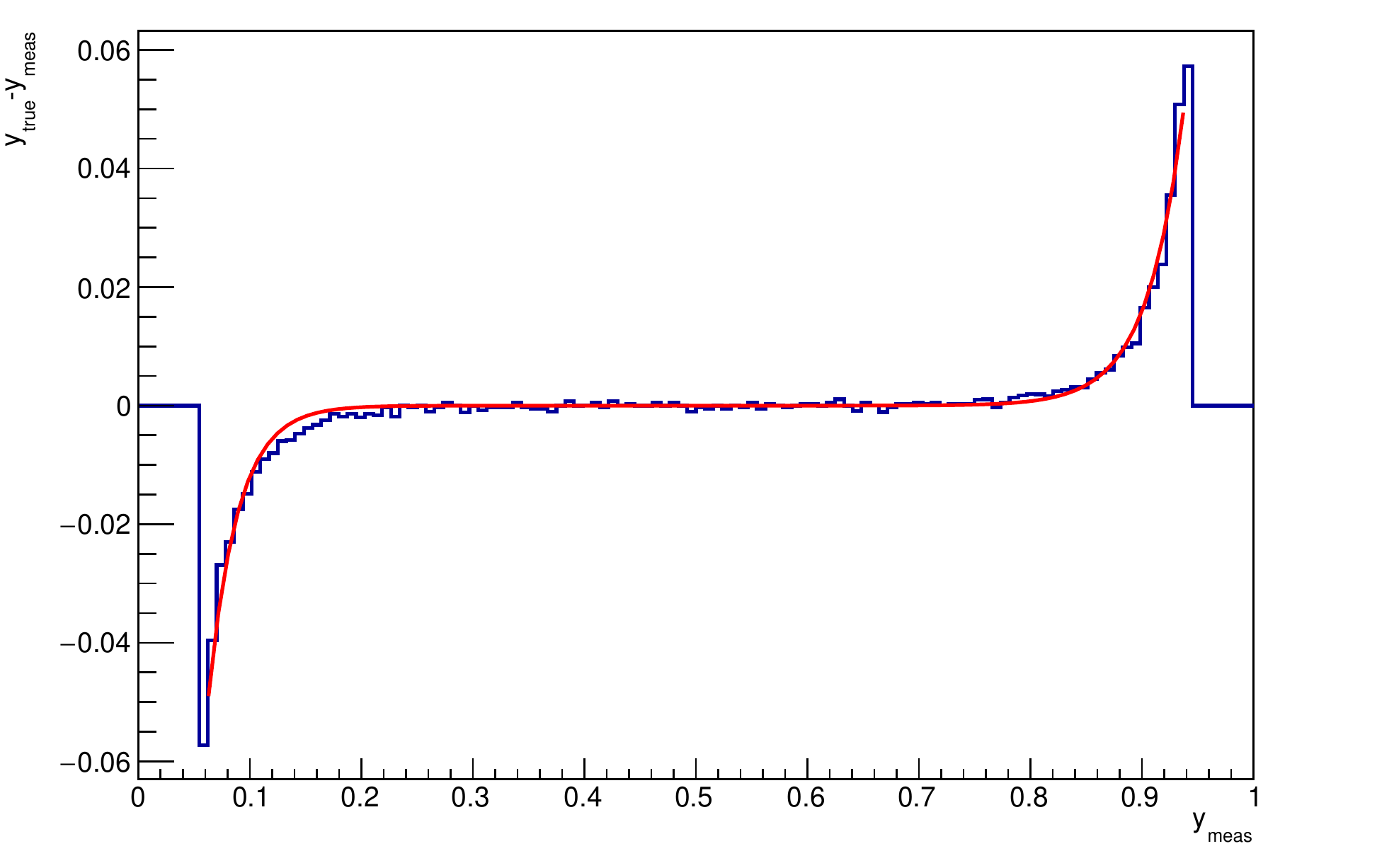}
  \caption{The edge bias $dy=y_{\mathrm{true}}-y_{\mathrm{meas}}$ is measured in the $y$ direction. The red line shows the best fit using the function in Eq.~\ref{eq:edgefit}.}
  \label{fig:edge_bias}
\end{figure}

To evaluate this bias along the $y$ direction, we move the source position horizontally, from $(y=0,z=0.5)$ to $(y=1,z=0.5)$. Using discrete positions along this axis, we measure the deviation between the measured and the true positions: $dy=y_{\mathrm{true}}-y_{\mathrm{meas}}$. This deviation is represented as a function of $y_{\mathrm{meas}}$ in Fig.~\ref{fig:edge_bias}. Usually, this bias can be mitigated using padding or windowing methods. Padding methods would increase data sizes and would lead to longer Fourier transform calculations. Windowing the data would alter the signal and noise counts used to compute the signal-to-noise ratio. Instead, the bias $dy(y_{\mathrm{meas}})$ is fitted by an empirical function:
\begin{equation}
  \eta_1(y_{\mathrm{meas}})=
  \begin{cases}
    H_0\times e^{H_1\times y_{\mathrm{meas}}} & \mathrm{if}\quad y_{\mathrm{meas}} > 0.5 \\
    H_2\times e^{H_3\times y_{\mathrm{meas}}} & \mathrm{otherwise},
  \end{cases}
  \label{eq:edgefit}
\end{equation}
where $H_0$, $H_1$, $H_2$, and $H_3$ are the fit parameters. The bias in the $z$ direction is corrected using the same method. Indeed the edge corrections in the $y$ and $z$ directions can be be addressed separately except for a small region near the corners of the camera plane where the corrections should not be decoupled. Given the low probability of finding a source in these positions, we neglect this effect. 

\begin{figure}
  \center
  \includegraphics[width=14cm]{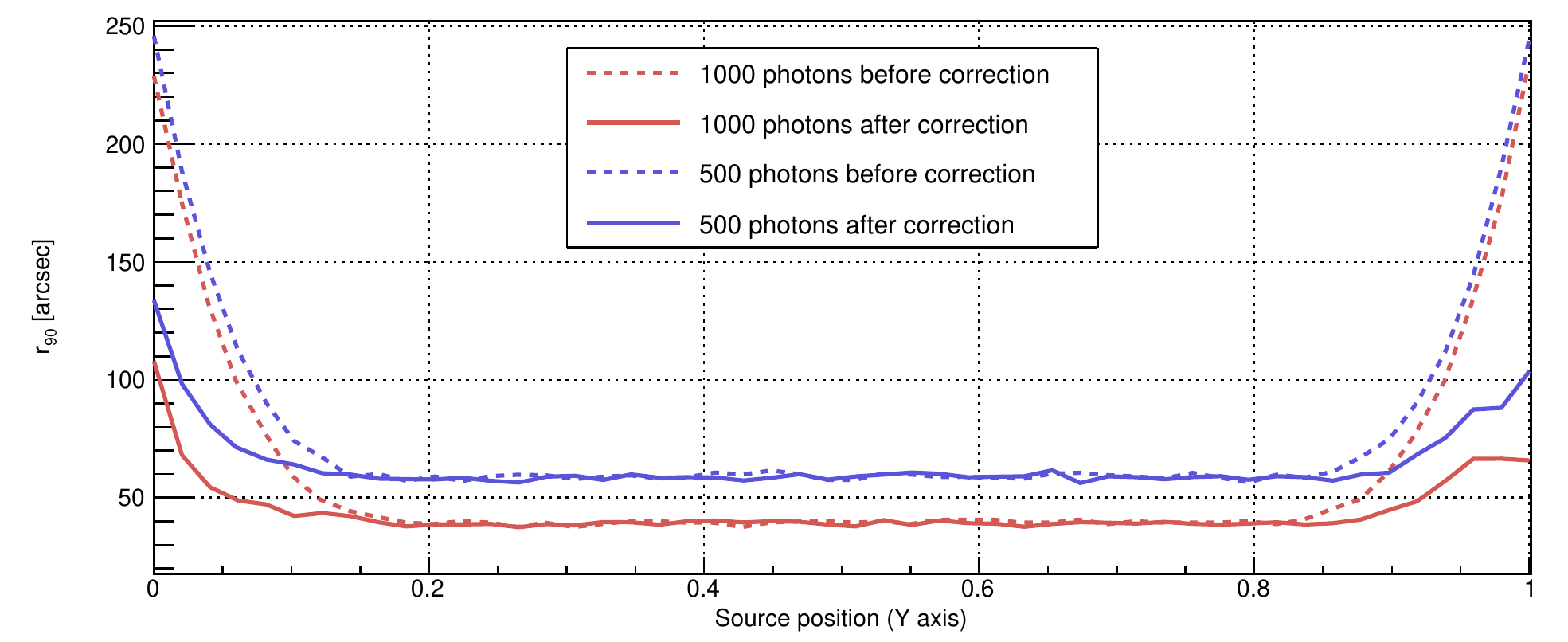}
  \caption{Localization uncertainty $r_{90}$ when the source position moves across the camera plane along the $y$ axis ($z=0.5$). The dashed lines do not include the edge-bias correction while the continuous lines do. The background count is fixed to 600 and two source intensities are tested: 1000 photons (red curves) and 500 photons (blue curves).}
  \label{fig:char_edge_bias}
\end{figure}

The $\eta_1$ and $\eta_2$ corrections are applied in Eq.~\ref{eq:peakposition_corrected} to obtain the final source position. Figure~\ref{fig:char_edge_bias} shows the localization uncertainty when the source position moves across from the camera plane with and without the edge-bias correction. The edge bias appears when the source is less than 30 pixels away from the camera edge. After correction, the bias is still visible but it is below the MXT design requirement of 2 arcmin. This comparison is done with 500 and 1000 photons showing that the edge correction method applies equally to all source intensities.

%%%%%%%%%%%%%%%%%%%%%%%%%%%%%%%%%%%%%%%%%%%%%%%%%%%%%%%%%%%%%%%%%%%%%%
\subsection{Multiple sources}\label{sec:characterization:multiple}
As explained in Sec.~\ref{sec:algorithm:multiple}, three X-ray sources are identified and localized in the field of view of the MXT. To characterize the algorithm with additional sources, we consider the worst-case scenario where two X-ray sources share the same $z$ coordinate; given the shape of the point spread function, the photon patterns of the two sources overlap. We fix the brightest source at center of the field of view and we move the second source from the center to the camera edge in the $y$ direction. We also vary the intensity of the second source relatively to the first source. The two sources are localized running the algorithm described in Sec.~\ref{sec:algorithm:multiple} and we measure the localization bias, $dr=\angle(\vec{r}_{\mathrm{meas}}, \vec{r}_{\mathrm{true}})$, for each source.

\begin{figure}
  \center
  \includegraphics[width=8cm]{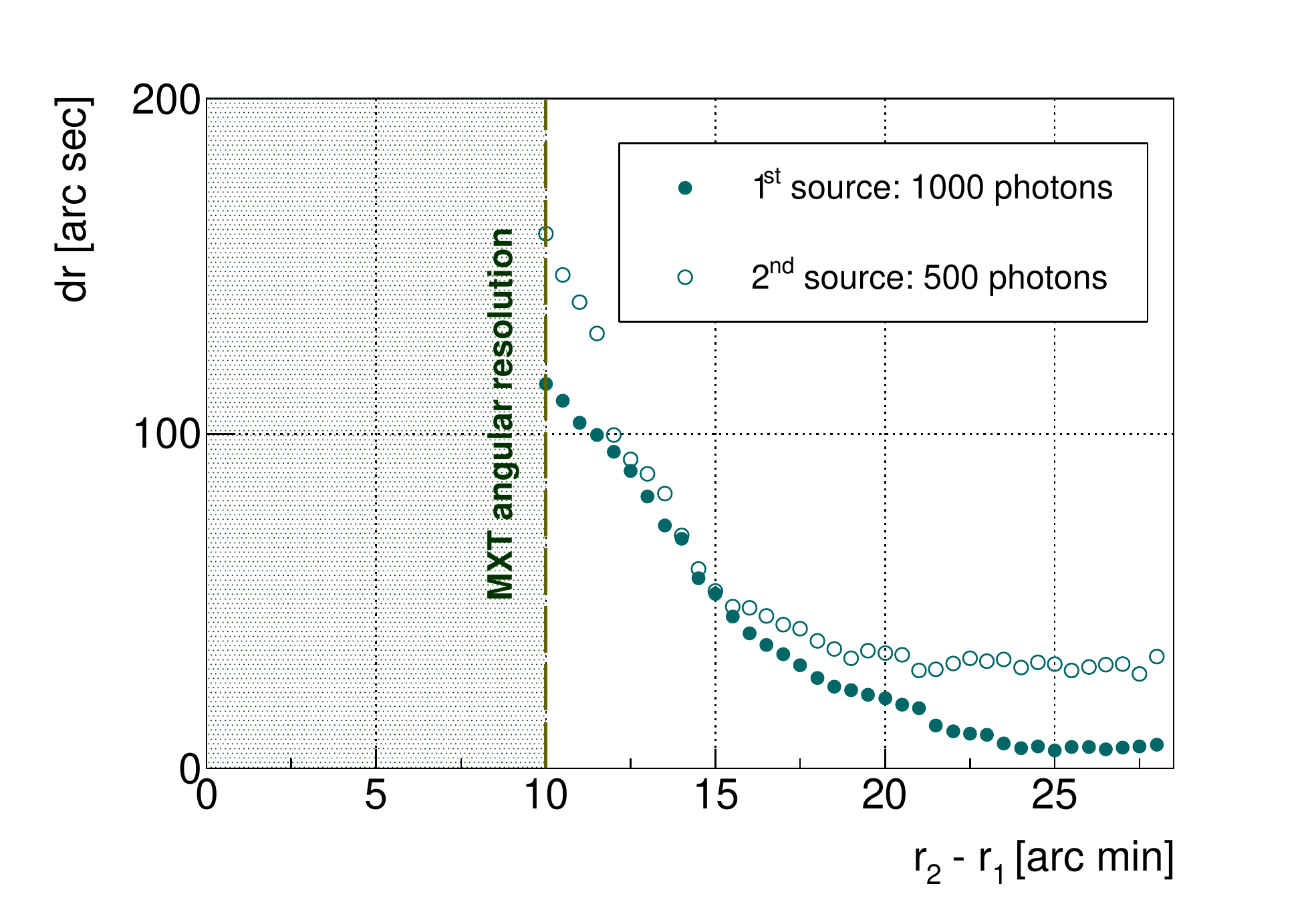}
  \includegraphics[width=8cm]{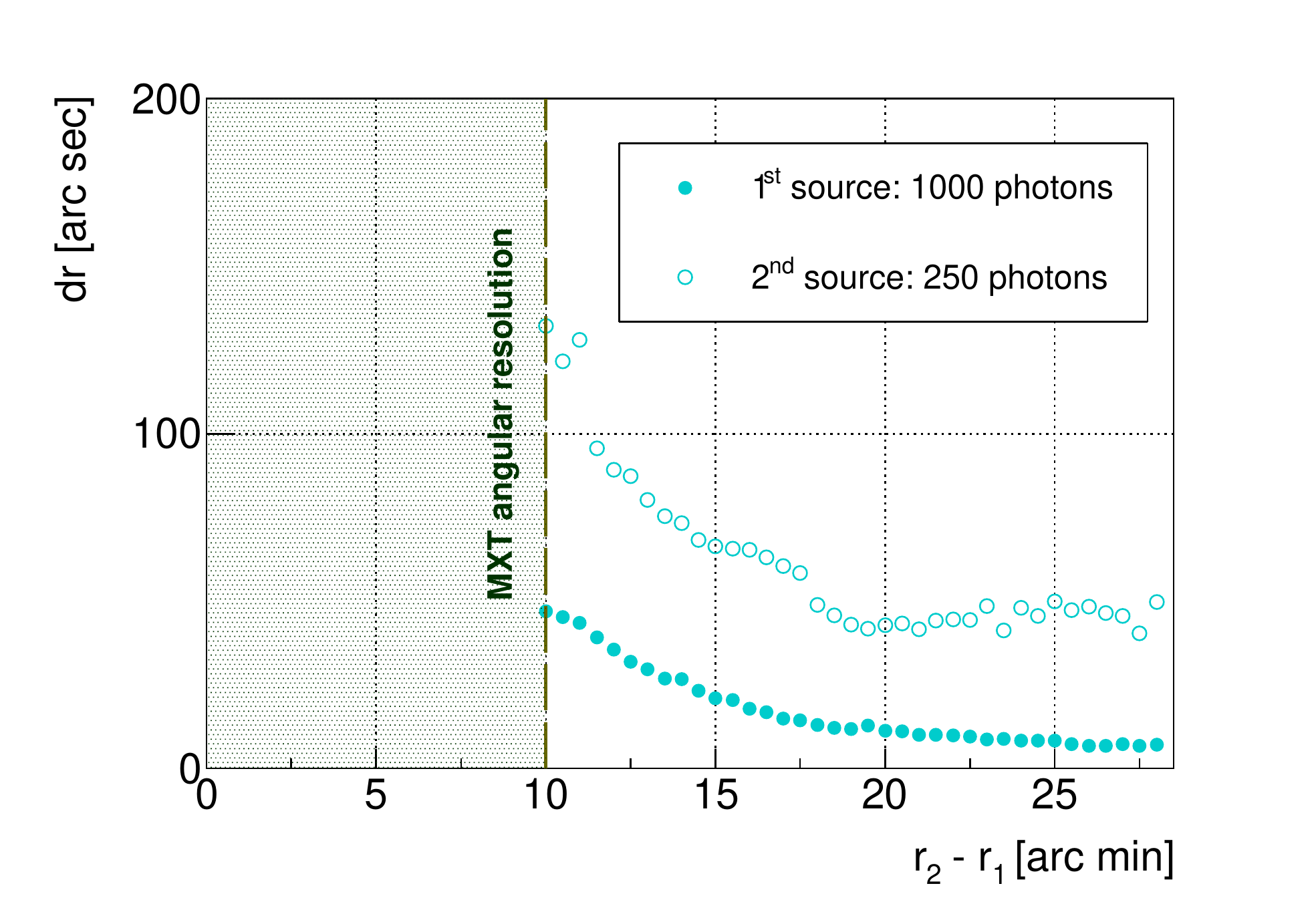}
  \caption{Localization bias ($dr$) when two co-aligned X-ray sources are present in the MXT field of view as a function of the angular distance between the two sources ($r_2-r_1$). We have masked the region where $r_2-r_1<10$~arcmin (in green) for which the peaks of the point spread function are overlapping and cannot be separated. The intensity of the first source is fixed to 1000 photons. Two intensities for the second source are tested: 500 photons (left plot) and 250 photons (right plot).}
  \label{fig:msources}
\end{figure}

Figure~\ref{fig:msources} shows the variation of $dr$ when the second source is moved away from the first source. The two sources cannot be separated as long as their angular distance is below 10~arcmin, which is approximately the size of the point spread function central peak, as expected. When the separation between the two sources is between 10 and 20~arcmin, the localization is biased by the presence of another source. Above 20~arcmin, the interference between the two sources can be neglected. We also test the impact of the relative intensity between the two sources. When the second source is less intense by a factor two, the localization performance for the first source is improved roughly by the same factor. The localization of the second source also improves thanks to a better subtraction of the first source.

%%%%%%%%%%%%%%%%%%%%%%%%%%%%%%%%%%%%%%%%%%%%%%%%%%%%%%%%%%%%%%%%%%%%%%
\subsection{Signal and noise counts}\label{sec:characterization:snr}
\begin{figure}
  \center
  \includegraphics[width=12cm]{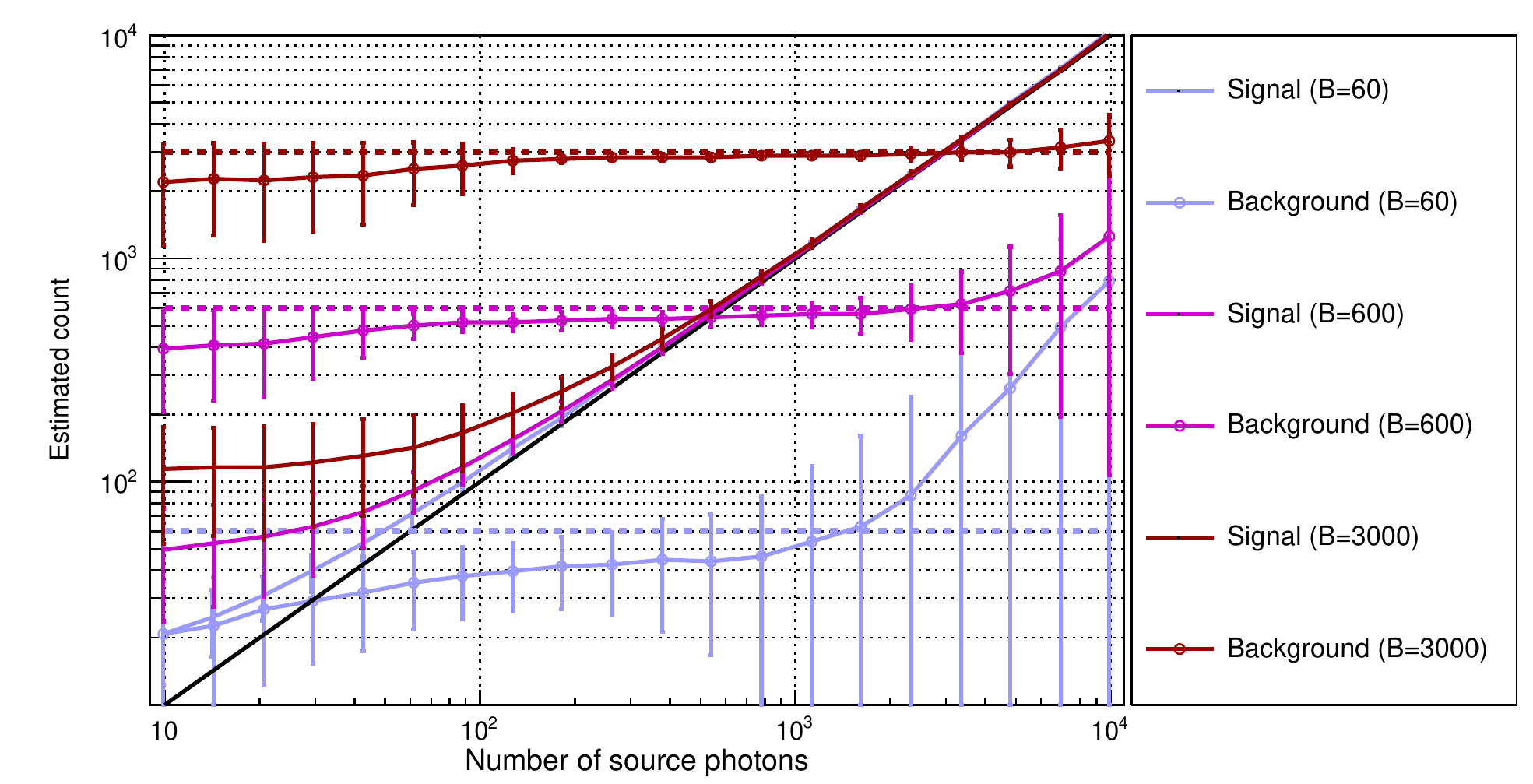} %
  \caption{The signal (smooth curves) and background (curves with markers) counts are estimated as a function of the signal intensity. For each point, 1000 source positions are randomly drawn in $0.3<y<0.7$ and $0.3<z<0.7$. Three background levels are considered: 60 counts (blue), 600 counts (magenta) and 3000 (dark red). They are represented by horizontal dashed lines.}
  \label{fig:sandb}
\end{figure}

In Sec.~\ref{sec:algorithm:snr}, we developed an inovative but non-trivial method to estimate the signal and background counts. To test the signal and background estimators, we vary the source intensity ($S_1$) and the background level ($B$). Both quantities are then estimated using Eqs.~\ref{eq:s1_estimate} and~\ref{eq:b_estimate}. The result of this study is presented in Fig.~\ref{fig:sandb}. The signal count is systematically over-estimated because of the simplification used to estimate the $C_{\varepsilon_1}$ matrix (see Sec.~\ref{sec:algorithm:snr}). For instance, when $S_1\simeq 200$, the signal is over-estinated by $\sim 10\%$ when $B=600$. The same effect leads to a systematic bias when estimating the background count: it is under-estimated for low signal fluxes and over-estimated for large signal fluxes.

\begin{figure}
  \center
  \includegraphics[width=10cm]{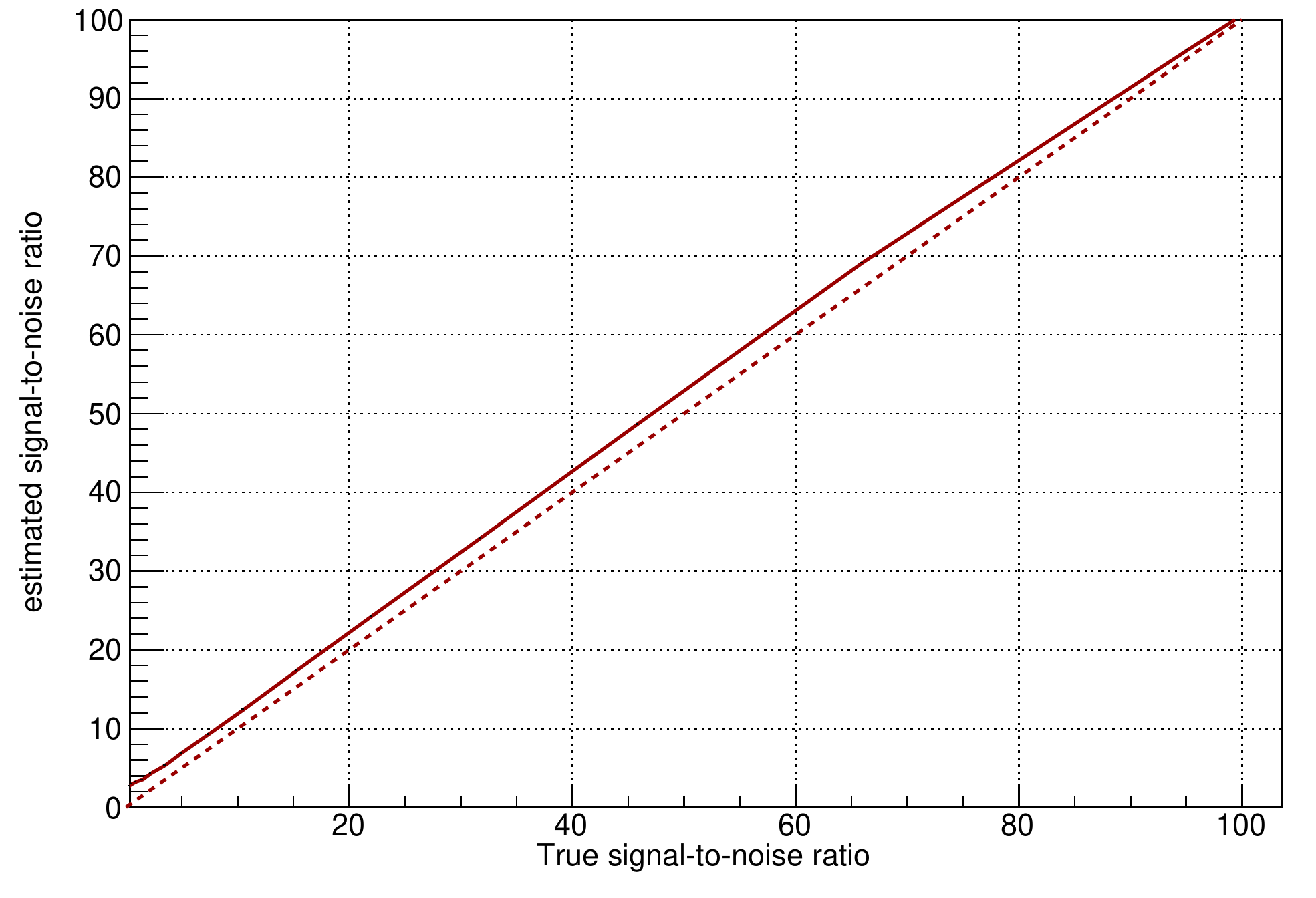} %
  \caption{Estimated signal-to-noise ratio as a function of the true signal-to-noise ratio, fixing $B=600$. The diagonal is indicated by a dashed line.}
  \label{fig:snr}
\end{figure}

Figure~\ref{fig:snr} shows the resulting signal-to-noise ratio, $\rho_1=S_1/\sqrt{B}$, estimated when fixing $B=600$. It is slightly over-estimated: for $10<\rho_1<100$, the signal-to-noise bias does not exceed 20\%.

The method to estimate $S$ and $B$ must be robust when several sources are present in the MXT field of view. The particular example of Fig.~\ref{fig:multiple}, and commented at the end of Sec.~\ref{sec:algorithm:multiple}, shows that our estimators give satifactory results. This method to estimate counts is iterative and is, therefore, intrinsically biased by the presence of less intense sources of X-ray photons. This bias is acceptable for on-board calculations. More advanced analyses can be conducted offline using X-ray source catalogs and multiple-source fitting techniques.

%% file: Conclusion.tex
\section{Summary}\label{sec:conclusion}
We presented the localization algorithm developed to process the images captured by the Microchannel X-ray Telescope mounted on the SVOM satellite. The images are analyzed on-board to rapidly detect and localize X-ray sources in the telescope field of view. Given the Lobster-Eye optical design of the telescope and the resulting cross-shaped structure of the point spread function, we developed specific analysis methods. Cross-correlation techniques were selected to maximize the sensitivity to faint sources. This paper shows that a localization accuracy better than 2~arcmin can be achieved when the telescope cumulates $\sim 150$ X-ray photons. This number guarantees that more than 85\% of gamma-ray bursts will be localized with an uncertainty better than 2~arcmin after 30 minutes~\citep{Gotz:inpress}. Moreover, we developed a new and reliable method to estimate the signal and background counts from the cross-correlation map from which we derive the signal-to-noise ratio for multiple sources in the field of view.

The MXT localization algorithm relies on many parameters, some of which were characterized and presented in this paper. All of them are configurable from the ground. During the first months of the SVOM mission, these parameters will be adjusted using real data collected in space. They will then be uploaded on board to maximize the performance of the localization algorithm. These parameters will be updated over time to follow aging effects. 

The SVOM satelite will be launched in 2023 and will observe gamma-ray bursts shortly after that. Moreover, SVOM will continue X-ray observations and step over when other satellites, such as Swift~\citep{Gehrels:2004qma} or Fermi~\citep{Michelson:2010zz}, will terminate their missions. SVOM will play an important role in the multi-messenger astronomy, in particular in coincidence with gravitational-wave detectors. The localization algorithm described in this paper will provide a rapid follow-up of gravitational-wave alerts to identify an electromagnetic counterpart.